# Green synthesis of eco-friendly silver nanoparticles using *Coffee arabica* leaf extract and development of a cost-effective biosensor for cysteine

*E. S. Harsha Haridas[1], Susmita Bhattacharya[2], M. K. Ravi Varma[1], Goutam Kumar Chandra*[\*,1]*

[1]Department of Physics, National Institute of Technology Calicut, Kozhikode-673601, Kerala, India.

[2]Radiation Laboratory, University of Notre Dame, Notre Dame, Indiana 46556, USA.

*Email: goutam@nitc.ac.in

KEYWORDS: *Coffee arabica* leaf; Phytosynthesis; Silver nanoparticles; Amino acids; Cysteine; Colorimetric sensing; Limit of detection; Raman spectroscopy

ABSTRACT:

The thirst for fabricating environment friendly silver nanoparticles (AgNPs) as biosensors via the green route is highly demanded among researchers due to their free of cost and less impact on creating pollution compared to other methods involving risk interference from toxic chemicals as reducing, stabilizing and capping agents. Here, for the first time, *Coffee arabica* leaf extracts (CAE) were used for the production of highly stable AgNPs in an easy and eco-friendly manner.

Within ten minutes, we can visually confirm the reduction of $Ag^+$ to $Ag^0$ from a change of yellow color to dark brown by the addition of silver nitrate ($AgNO_3$) solution with CAE. The structural, optical, stability, morphological and elemental characteristics of CAE-reduced AgNPs (CAE-AgNPs) were investigated using ultraviolet-visible (UV-Vis) spectroscopy, Fourier transforms infrared spectroscopy in ATR mode (FTIR-ATR), Raman spectroscopy, Zeta potential analyzer and transmission electron microscopy (TEM). The prepared CAE-AgNPs show good sensing performance for the amino acid L-cysteine (Cys) with high sensitivity, a limit of detection (LOD) of 0.1 nM, selectivity and good stability. Hence, the proposed novel method is highly promoted among other nanoengineered green AgNPs in the medical field as a valuable biosensor.

**Introduction**

The usage of silver nanoparticles (AgNPs) for various consumer goods like health care products, sensing devices, cosmetics, food and engineered devices is increasing, due to their unique physical and biological properties[1-3]. Synthesis of stable AgNPs with proper distribution and properties as well as biocompatibility is important and efficient for their activities. Various methods have been reported for producing AgNPs, including chemical reduction, evaporation, laser ablation, condensation, the irradiance of laser, microwave or electrons, ionization, and photochemical routes and green synthesis using either microorganisms or plant parts[4-7]. Except for those green synthesis routes, most existing methods use toxic chemicals as reducing and capping agents and pose environmental problems. Some are time consuming and require controlled conditions, complicated procedures, experienced technicians, proper hygiene and heavy investments. Moreover, currently available AgNPs are in powder mode, hence studies involving aqueous re-dispersion of AgNPs is unable to proceed further because re-dispersion affects the nature of the prepared AgNPs.

In this context, the green chemistry approaches play a vital role to implement sustainable and environmental friendly pathways to produce desirable modes with uncontaminated products having minimum toxicity and side effects. The green reduction method especially using plant parts, commonly known as "phytosynthesis", is an alternate synthesis procedure rather than using microbial that involve biological risk factors. There are several advantages of phytosynthesis routes over other methods like, it is easy and simple to conduct, safe to handle, zero-energy based, readily available, not require controlled ambience, rapid and single-step process, cheap, renewable, and no utilization of harsh and costly chemicals, zero waste products and finally eco-friendly[7].

Fourier transform infrared (FTIR) spectroscopic studies reveal that the primary phytochemicals present in plants like alkaloids, polyphenols, terpenoids, glycosides etc., involves in the reduction of Ag ions ($Ag^+$) to AgNPs due to metal ion hyper accumulating and reductive capacity[8]. Coffee arabica plant is a better choice for green reduction of $Ag^+$ because of its high antioxidant capacity[8-10]. While trimming the branches of coffee plants, the leaves are usually discarded in the field and considered low- or no-value by products. The phytochemicals present in coffee leaves are excellent reductants, capping agents and stabilizers like caffeic acids (CA), chlorogenic acids (CGA), mangiferin, carbohydrates, amino acids and rutins that can be used for green synthesis of NPs. But it is being neglected due to the high preference placed on seasonal and costly coffee beans. Recently, one research group have studied the structural properties and yield of ZnO NPs using the coffee leaf extract as a reducing agent[11]. To our best knowledge, there is no study using coffee arabica leaf extracts (CAE) to synthesize AgNPs. So, it would be interesting to synthesize CAE-reduced AgNPs (CAE-AgNPs) to produce value-added products and apply them to various fields.

Green synthesized AgNPs are being used in various molecular diagnostic processes like bio sensing[12], bioimaging[13], etc. due to their simplicity, less toxicity, high sensitivity, specificity and cost-effectiveness. The sensing ability of AgNPs is enhanced via surface plasmon resonance (SPR), spectrophotometric emission and absorption. Since AgNPs show a high affinity towards nitrogen and sulfur-containing molecules[14], they can be applied in the field of low-level precise detection of biomolecules[15]. This is one of the reasons for the increasing demand of developing AgNPs based biosensors for L-cysteine (Cys) which plays a crucial role to track biological pathways. Cys, the building block of proteins belonging to non-essential amino acids, is polar and uncharged and essential for cellular systems in living things. It controls detoxification, helps in protein metabolism and suppresses the ageing of the skin by preserving its texture and elasticity by producing collagen. The imbalance in the levels of Cys in biofluids induces health disorders like cystinuria, hair loss, and growth retardation, fat loss, liver damage, white blood cell loss, etc.[16-18] Hence, an efficient sensing method for Cys at trace amounts is vital for pre-diagnosing disorders.

There have been various selective and sensitive detection methods for Cys monitoring which include fluorescence analysis[19], electrochemical method[20], bioluminescence[21], high-performance liquid chromatography (HPLC)[22], voltammetry[23], spectroscopic method[24-25] etc. The majority of these techniques require specific chromophores for specific analytes, various analytes pre-treatment steps, highly expensive instruments, high energy consumption and use of a large number of synthetic solvents. On the other hand, some methods of Cys detection suffer disadvantages due to long reaction time, limits in detection, involvement of highly sophisticated machinery, and technicians as well as suitable physical conditions and high investment. So, compared with other techniques colorimetric detection using green reduced AgNPs is simple, environmentally friendly, rapid and cost-effective.

In this study, we have introduced the green synthesis of AgNPs using CAE for the first time and it opens a new domain in the field of fabrication of highly stable AgNPs for a sustainable environment. To explore the practical value of CAE-AgNPs, we have reported its sensing ability towards Cys and the interaction of the analyte with AgNPs is discussed using vibrational assignments in Raman and FTIR spectroscopic techniques.

**Experimental**

*Materials*

Fresh and young *Coffee arabica* leaves were collected from the local coffee plantation of Vythiri village, Kerala, India. The leaves were washed several times with deionized (DI) water to remove the dust. Silver nitrate ($AgNO_3$) with a 98% purity and Cys were purchased from Sigma-Aldrich Chemical Co. L-Alanine (Ala), L-Arginine (Arg), Butyric acid (Ba), L-Glutamic acid (Glu),

Glycine (Gly), L-Isoleucine (Ile), L-Tryptophan (Trp) and L-Valine (Val) were purchased from LobaChemie Pvt. Ltd. All chemicals were used without any further purification. All experiments were carried out using DI water.

*Preparation of CAE*

10 gm of washed coffee leaves (Scheme 1) were cut into small pieces and boiled with 250 ml of DI water at 100 °C for 5 min. The extract was allowed to cool up to room temperature. The obtained yellow color extract was filtered using Whatman No.1 filter paper and the supernatant was stored at 4 °C for further studies.

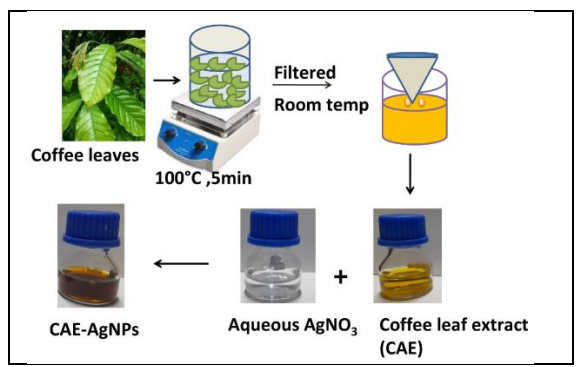

**Scheme 1.** Schematic of the synthesis procedure for CAE-AgNPs.

*Preparation of CAE-AgNPs*

The AgNP colloid was prepared by adding 1 ml of CAE to 5 ml aqueous solution of $AgNO_3$ (1 mM) followed by gentle shaking at room temperature. Within 5-10 min the color of the solution changed from pale yellow to brownish-yellow, which indicates the reduction of $Ag^+$ to AgNPs. The CAE-AgNPs suspension was also collected for different $AgNO_3$ concentrations (1, 2, 3 and 4 mM) and CAE volumes (2, 3, 4 and 5 ml), respectively (Fig. 1b and 1c).

Characterization

The characterization of green synthesized AgNPs was done using a UV-1800 Shimadzu double beam spectrophotometer of 1 nm resolution in the wavelength range 200 - 800 nm. A quartz cuvette of path length 1 cm was used for collecting the absorption spectrum. The bio-reduction of $Ag^+$ in aqueous plant extract at various concentrations of CAE and metal ions was monitored by diluting the suspension using DI water. The size, morphology and crystalline nature of sample A1 (CAE-AgNPs suspension with 1 ml of CAE) before and after adding an adequate amount of Cys was investigated by high-resolution transmission electron microscopy (HR-TEM). TEM micrographs were taken using FEI TecnaiG2 Spirit Bio-Twin TEM at an accelerating voltage of 300 kV and an ultra-high-resolution pole piece. The samples were prepared by placing a drop on a graphite grid and drying it in a vacuum. Surface analysis of CAE and CAE-AgNPs was done using PerkinElmer frontier FT-IR spectrometer of Attenuated total reflection mode (FTIR-ATR) and Confocal Raman Microscope with 785 nm laser (Horiba France SAS Lab RAM HR Evolution). A particle size analyzer of Malvern Nano ZS (4 mW, 633 nm) was used for measuring surface load and the average size of A1. Stability studies were done by checking the zeta potential of AgNPs prepared using different volumes (1, 2, 3, 4, and 5 ml respectively) of CAE by Zeta Check-Particle Charge Reader (Microtrac, PMX 500).

Selective detection of Cys

For the selectivity study, 1 ml of each of the nine different amino acids (Ala, Ba, Gly, Cys, Arg, Trp, Gla, Va and Isl) with 1 mM concentration were added to A1 and the color change and absorbance were noticed under the same conditions as that of the sensitivity study using UV-Vis measurements.

Sensitive detection of Cys

A1 was diluted 50 times and the pH was adjusted to 6.0. 1 ml of Cys with different concentrations ($2\times10^{-3}$ to $2\times10^{-8}$ M) was added to 1 ml of A1. After 5 minutes the color change was observed and the corresponding SPR bands were studied by UV-Vis measurements. To find out the limit of detection (LOD) using Raman measurements, additionally, two more diluted Cys samples were used.

**Results and discussion**

Characterization of CAE and CAE-AgNPs

Caffeine and CGA are the most powerful purine alkaloid and phenolic components found abundant in coffee leaves in the ratio of 1:4 (per mg/g of dry matter) which is the reason behind the popularity of coffee as a stimulant[26]. The presence of caffeine and CGA is evident from the peak centered at around 274 nm and a small hump at a higher wavelength range[27] in the UV-Vis absorption spectrum (Fig. 1a). These regions can be attributed to the $n \rightarrow \pi^*$ electronic transition of caffeine and CGA, respectively[28].

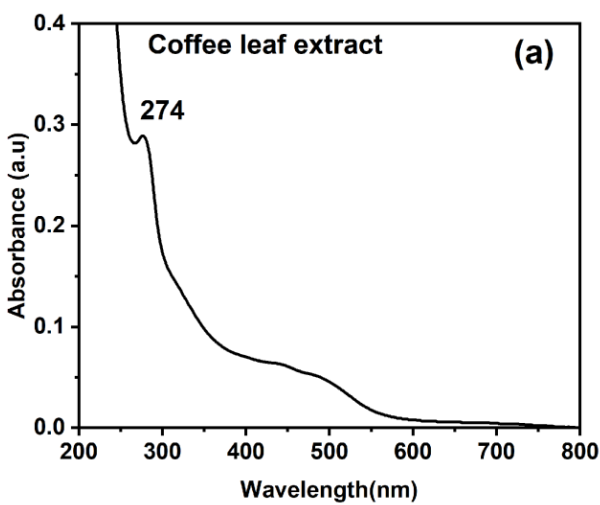
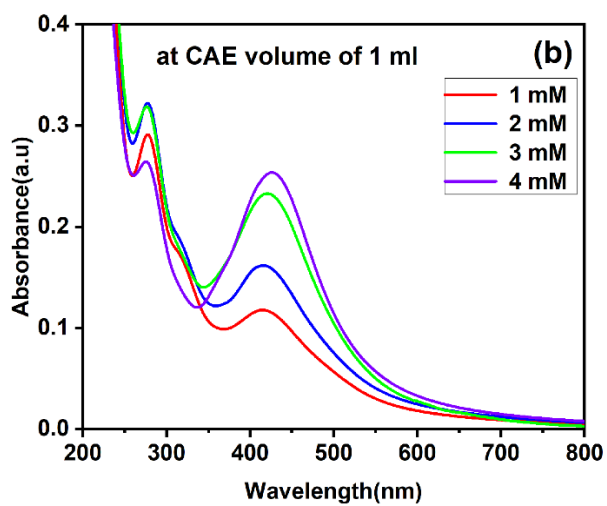

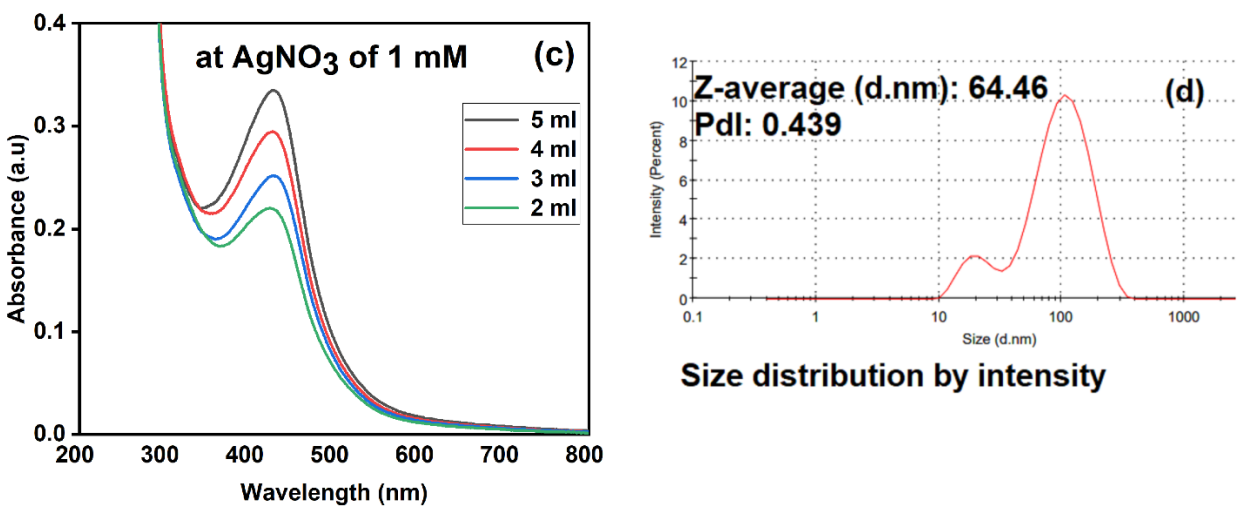

Figure 1. UV-Vis spectrum for (a) aqueous solution of CAE (b) CAE-AgNPs formed using various $AgNO_3$ concentrations; (c) CAE-AgNPs formed using various CAE volumes. (d) DLS spectrum of CAE-AgNPs.

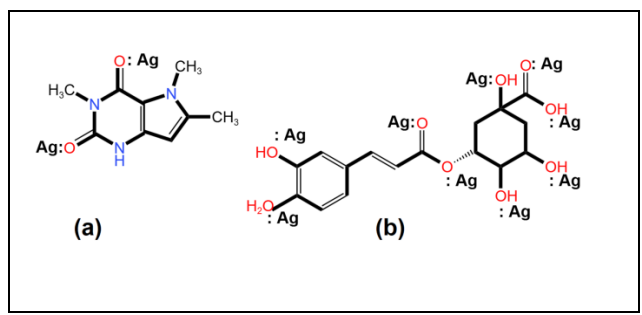

Figure 2. Proposed binding scheme of the major phytochemicals (a) caffeine and (b) CGA of CAE, attaches on AgNPs via nucleophilic oxygen atoms of respective carbonyle and hydroxyl groups.

Hence, the reduction of $Ag^+$ is plausibly due to caffeine and other CGA. At alkaline pH there are two possibilities of binding of caffeine on $Ag^+$, either through one of the oxygen atoms or through nitrogen[29]. On the other hand, CGA exists in $COO^-$ form at pH above 5.0. Thus, for CGA it can be bound to Ag directly via $COO^-$ or through $O^-$ group[30] (Fig. 2). Generally, noble metal NPs like AgNPs exhibit $SPR^2$, a collective oscillation of the conduction electrons on the metal surface when they are excited by light of proper wavelengths. Thus, CAE-AgNPs synthesis is monitored using a UV-Vis spectrometer and the corresponding effect of concentration of CAE and $AgNO_3$ on the formation of CAE-AgNPs is studied and shown in Fig. 2b – c. As $Ag^+$ concentration increases from 1 to 4 mM, a red shift is occurred in the absorption band along with an increase in intensity (Fig. 1b). This might be due to the formation of a large number of AgNPs with an increase in size[31]. On the other hand, a slightly blue shift is observed in absorption peak when the volume of CAE varies from 2 to 5 ml indicating the reduction in particle size[31]. In addition, a decrease in the full-width half maxima (FWHM) of the absorption peak occurs as the volume of the reducing agent is increased. This may be due to the donation of electron density from the surface of AgNPs to CAE[7]. These results indicate that the synthesized AgNPs using 1 ml of CAE at 3 mM $AgNO_3$ concentration are the finest and more stable with a zeta potential value of -57 eV even after three months of synthesis. The high value of zeta potential indicates the excellent stability of as-prepared samples. Thus, this AgNPs is selected for further analysis like TEM, FTIR, Raman spectroscopy, DLS, respectively and Cys detection.

Morphological studies of CAE-AgNPs were investigated using TEM which shows that particles are dispersed well and spherical symmetry is predominant (Fig. 3a - c). HRTEM images reveal the poly crystallinity of CAE-AgNPs with a clear lattice fringe with the spacing of 0.26 nm corresponding to Ag (111) plane of face-centred cubic (fcc) lattice[32] (Fig. 3d).

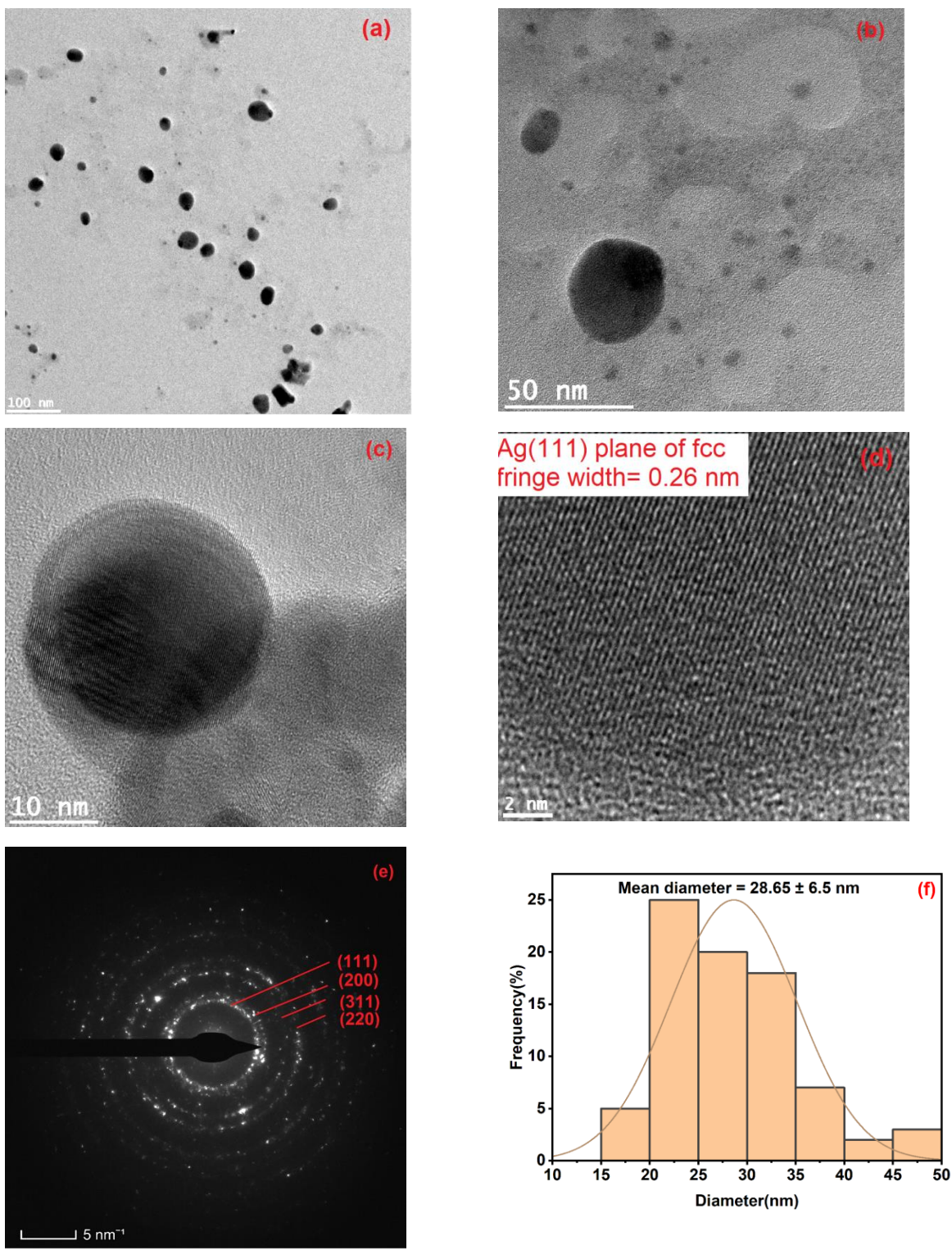

**Figure 3.** TEM images of CAE-AgNPs at different magnifications (a-c), lattice fringe (d), SAED pattern (e) and histogram for size distribution (f).

The selected area electron diffraction (SAED) pattern (Fig. 3e) confirms the (111), (200), (220) and (311) planes of AgNPs[32]. The average size of CAE-AgNPs estimated from TEM is 28.65 ± 6.5 nm (Fig. 3f). From the DLS measurement (Fig. 2d) the average hydrodynamic diameter of CAE-AgNPs was determined and is 64.46 nm which is larger than that of TEM. In DLS, the hydrodynamic diameter gives us information on the phytochemical cloud around the NPs as it moves under the influence of Brownian motion. In TEM while estimating the size of the particle the hydrodynamic core corresponding to the surfactant is absent, there we are measuring the projected area diameter[33].

| Sample | Wavenumber (cm$^{-1}$) | IR Assignments | Reference |
|---|---|---|---|
| CAE | 1299 | Vibrations of polyols | 38 |
| | 1640 | C=O stretching vibration | 34,38 |
| | 2140 | C-H vibrations | 34 |
| | 3291 | O-H stretch | 35,38 |
| CAE-AgNPs | 478, 590 | Ag-O, Ag-N vibrations | 35,38 |
| | 1040, 1238 | Vibrations of primary alcohol (-OH) and polyols | 34, 38 |
| | 1353 | Aromatic band vibrations | 35 |
| | 1640 | C=O, C=N stretch | 34 |
| | 3306 | O-H stretch | 35,38 |

**Table 1.** FTIR vibrational mode assignment for CAE and CAE-AgNPs.

FTIR-ATR technique, a vibrational spectroscopic tool is useful for the qualitative and quantitative analyses of materials due to its functional group specificity. The FTIR spectrum of CAE (Fig. 4a) exhibits characteristic bands (as tabulated in Table.1) at 3291, 2140, 1640 and 1299 cm$^{-1}$. The reduction, capping and stabilization process causes some significant changes in the FTIR spectra of CAE-AgNPs (Fig. 4b) like change in the spectral intensity and/or shift in the position of the peaks appear at 3306, 2117, 1640, 1353 and 1238 cm$^{-1}$ compared to the CAE. Some additional peaks are also emerged in CAE-AgNPs (Fig. 4b) and are situated at around 1785, 1238, 1040, 590 and 478 cm$^{-1}$. These spectral assignments signify the potential role of phytochemicals present in CAE. Here, the strong broadband centered at around 3291 cm$^{-1}$ and week band at around 2140 cm$^{-1}$ are ascribed to the stretching vibrations and first overtone of hydroxyl groups (-OH) of phenolic acids, flavanoids, xanthones and terpenes of CAE[34]. In case of CAE-AgNPs, these bands are shifted to 3306 and 2117 cm$^{-1}$, which reflects the crucial role of –OH groups towards reducing Ag$^+$ to AgNPs.

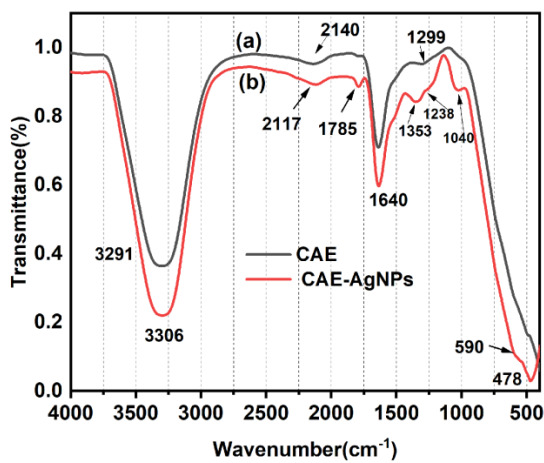

**Figure 4**. FTIR-ATR spectra in the region 4000-400 cm$^{-1}$ of CAE (40 mg/ml) and CAE-AgNPs (3 mM of AgNO$_3$ with 1 ml CAE) are indicated by black and red solid lines, respectively.

The emergence and more pronounced absorbance value of the carbonyl band at 1785 cm$^{-1}$ in Fig. 4b signifies the role of methyl esters, carbohydrates and lipids present in caffeine during the reduction process of formation of AgNPs[34]. The increased signal intensity of peak lies at 1640 cm$^{-1}$ denotes the out-of-phase C=O stretching vibrations mixed up with C=C stretching mode in caffeine[34]. The bands in the region 1300 - 1000 cm$^{-1}$ are assigned to the presence of CGA and also these are attributed to the polysaccharides and other carbohydrates[34]. The stretching modes of Ag-O can be seen in the wavenumber regions 590 and 478 cm$^{-1}$ which confirms the reduction of Ag$^+$ to AgNPs[35]. Hence, the role of phytochemicals presents in CAE mainly caffeine and CGA in the Ag$^+$ reduction mechanism is clearly suggested from the FTIR-ATR data.

| Sample | Wavenumber (cm$^{-1}$) | Raman Assignments | Reference |
|---|---|---|---|
| CAE | 450, 488 | Out of plane bending vibrations of -OH | 37 |
| | 600, 800 | C=C vibrations of alkenes<br><br>Out of plane bending vibrations of -C=O and -CH | 34,36,37 |
| | 1640 | C=O, C=N stretching vibrations | 34 |
| CAE-AgNPs | 230, 450, 488 | Ag-O, Ag-N vibrations<br><br>Out of plane bending vibrations of -OH | 38 |

|  | 600, 800 | C=C vibrations of alkenes<br><br>Out of plane bending vibrations of -C=O and -CH | 34, 36,37 |
|  | 1353 | Aromatic band vibrations | 35 |
|  | 1640 | C=O, C=N stretching vibrations | 34 |

**Table 2.** Raman vibrational mode assignment for CAE and CAE-AgNPs.

Raman spectroscopy is a simple and reliable tool that provides a detailed fingerprinting of molecules in a non-destructive and rapid manner. Fig. 5a-b shows Raman spectra of CAE and CAE-AgNPs in the range of 200 - 1800 cm$^{-1}$. For CAE, the major Raman vibrational modes are present at 450, 490, 600, 800 and 1640 cm$^{-1}$. Similarly, for CAE-AgNPs, a new band appears at 230 cm$^{-1}$ along with the other parent bands which show some shift in positions and variation in intensities. The vibrational modes and their assignment are tabulated in Table 2. Ag-O bending vibrations[35] is evident from the newly emerged band at 230 cm$^{-1}$ in Fig. 5b. The increased intensity of the peaks after the formation of AgNPs located at 450, 488 and 600 cm$^{-1}$ in Fig. 5b compared to CAE (Fig. 5a) corresponds to out-of-plane bending vibrations of -OH and -C=O groups present in the phenolic contents of CAE and indicates the involvement of those phytochemical in the reduction mechanism of Ag ions[36,37]. Moreover, the vibrational assignments due to out of plane bending vibrations of -CH and stretching of C=C of alkenes are also present at around the wavenumber regions 600 and 800 cm$^{-1}$. These bands are characteristic bands of caffeine present in CAE[34,36,37]. Hence, the role of phytochemicals present in CAE as reducing and capping agent during the formation of NPs is evident from the Raman spectral investigations of CAE and CAE-AgNPs, respectively.

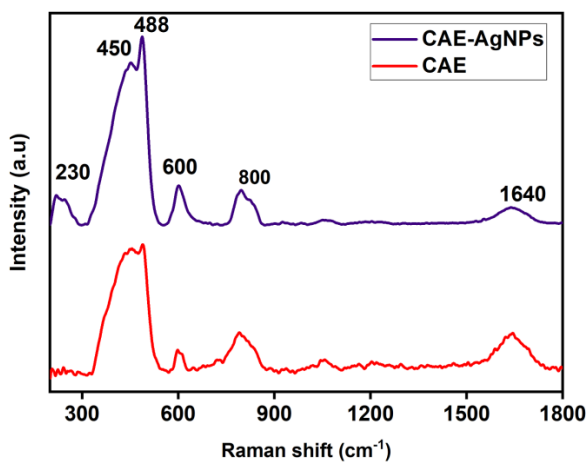

**Figure 5.** Raman spectra of (red) CAE (40 mg/ml) and (violet) CAE-AgNPs (3 mM of AgNO$_3$ with 1 ml CAE) in the region of 200-1800 cm$^{-1}$.

Selectivity and sensitivity studies of AgNPs for the detection of Cys

The interaction studies of nine different amino acids Ala, Ba, Gly, Cys, Arg, Trp, Gla, Val and Ils with CAE-AgNPs were carried out and corresponding colorimetric responses are noted (Fig. 6a). From Fig. 6a, it is clear that only the sample of AgNPs with Cys shows distinct color of dark reddish brown instead of pale yellow as in the case of others. The outcome indicates that synthesized CAE-AgNPs possess high selectivity towards the Cys sample than others. There are notably few spectral changes that occurred in the SPR absorption of AgNPs with the addition of other amino acids (Fig. 6a). In the case of Cys, the intensity of parent band of CAE-AgNPs drastically reduced and at the same time there is an appearance of new peak at the higher wavelength around 560 nm. It is mainly due to the chemical interaction between CAE-AgNPs and the analyte Cys which is followed by the aggregation of the particles that is confirmed from the TEM images shown in Fig.6c – d. The corresponding size of the NPs was 48 ± 3.2 nm and from the former spherical shape, the morphology is modified into an oval.

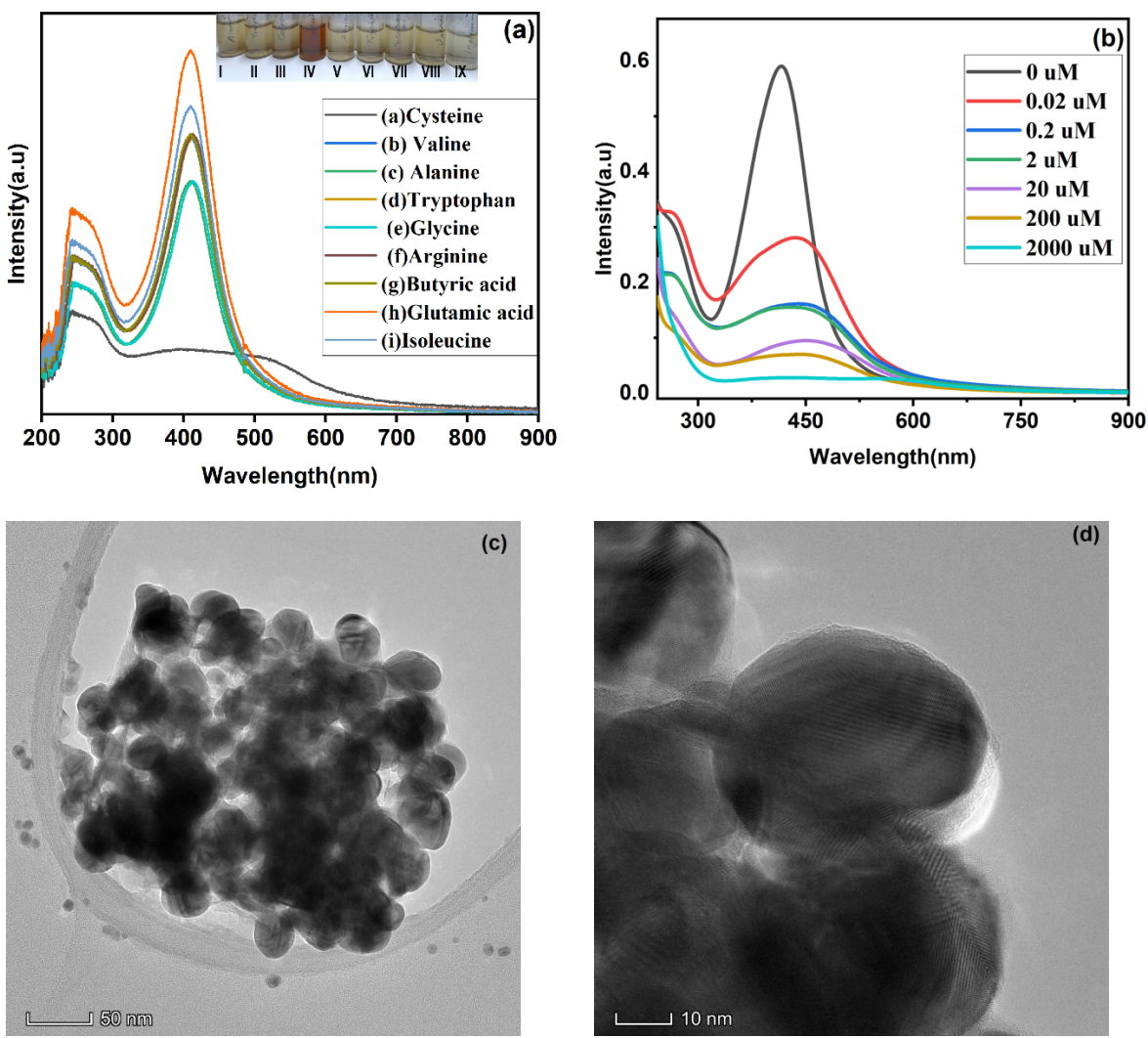

**Figure 6.** UV-Vis spectra of (a) selectivity study using different amino acids and the inset show color noticed for CAE-Ag colloid after the addition of nine amino acids (I- Ala, II- Ba, III- Gly, IV- Cys, V- Arg, VI-Trp, VII- Ga, VIII-Val and IX- Ils), respectively. (b) sensitivity study of Cys at different dilutions and (c) - (d) TEM images of CAE-AgNPs after adding Cys.

The increment in the average diameter is about 21.50 nm, indicating the aggregation of CAE-AgNPs. Further, an aqueous solution of Cys of various concentrations ranging from 0.02 μM - 2 mM was used to test the sensitivity (Fig. 6b). The intensity of the plasmon band of CAE-AgNPs centered at 414 nm decreases drastically and continually as the concentration of Cys is increased from 0.02 μM - 2 mM. At the same time, there is an appearance of new peak at the higher wavelength around 450 nm. The trend of decreasing intensity and appearance of new peak at around 450 nm can be attributed to the affinity of AgNPs to make bonds with the analytes- Cys present as surface ligands which leads to the aggregation[36].

*The mechanism for detection of Cys using CAE-AgNPs*

Being a low-cost colorimetric probe, AgNPs are becoming a promising sensing tool among researchers. Mainly their LSPR property is utilized for fabrication of LSPR-based nanosensors. The parameters like size, shape, inter particle distance, dielectric constant of the medium, surface characteristics and composition are in turn related to the LSPR phenomena. Here, the reason for distinct color change and drastically decreased absorption intensity of CAE-AgNPs after the addition of Cys (as shown in Fig. 6a - b) can be attributed to the adsorption of Cys on AgNPs surface via the plasmon-driven chemical transformations at the interface between AgNPs and Cys[36]. Hence, the approach of aggregation-based LSPR colorimetric sensing of CAE-AgNPs results in the reduction of inter-particle distance so that the adjacent particles start to overlap (Fig. 6c - d). Since AgNPs possess a high value of work function (4.26 - 4.75 eV), the binding of the NPs with electron rich groups are much more favorable[38]. So, the increased number density of hotspots of CAE-AgNPs makes a transformation in the chemistry of such NPs via energy transfer and hot electron carriers and enhances the reaction between Ag and Cys for the formation of Ag-Cys metal-ligand complex[36,39]. The active organic functional sites (-COOH, -$NH_3$ and -SH) of Cys act as ligating adsorbing species towards the AgNPs and allows to open up a visual pathway for analyzing metal species with enhanced selectivity and sensitivity[7]. The aggregation induced by Cys in the colloidal solution of AgNPs can be explained by the chemical interaction between the surface of AgNPs and thiol group present in the Cys. The isoelectric point of Cys is 5.02 and the reaction of Cys with CAE-AgNP is processing at a pH range of 6.3 - 6.7. So, the amino part of zwitterionic Cys will be neutralized and leaves an extra negative charge on carboxyl tail ($COO^-$)[7]. But compare to carboxyl group, the greater affinity of thiol group towards Ag metal NPs induces a replacement of CAE in CAE-AgNPs and gets adsorbed on the surface of AgNPs through ligand exchange reaction.

Since FTIR-ATR fails to study below the wavenumber region 400 $cm^{-1}$ at which metal-S bonds are usually formed, a far better choice is the Raman study whose signals are very susceptible to intermolecular interactions. Here, Fig. 7a - b shows the Raman spectra for pure Cys and CAE-AgNPs after the addition of Cys. Usually, Cys molecules forms disulfide bond (S-S) with adjacent Cys molecules and results in cystine. Here, the increased intensity of Raman signal at 488 $cm^{-1}$ in CAE-AgNPs after the addition of Cys (Fig. 7b) compared to CAE-AgNPs without Cys (Fig. 5b) can be attributed to the involvement of S-S bonding. On the other hand, the absence of the strong S-H stretching vibration of Cys at around 2550 $cm^{-1}$ in CAE-AgNPs indicates the confirmation of binding of sulfur part from Cys to AgNPs surface after the cleavage of thiol S-H group[40,41]. Hence, Ag-S bindings are confirmed in the region 200 – 500 $cm^{-1}$ and strengthen the possible evidence for greater affinity of CAE-AgNPs towards sulphur ions.

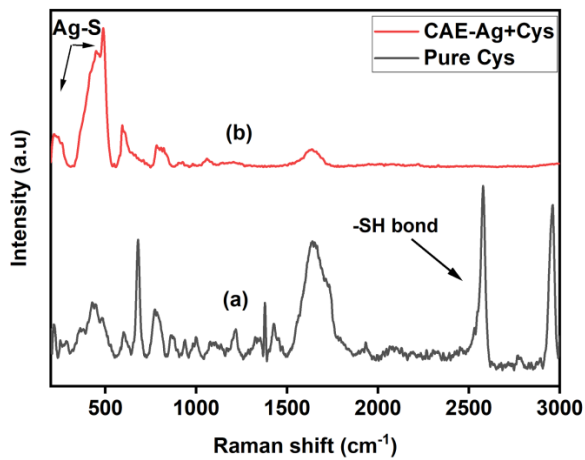

**Figure 7.** Raman spectra of (a) pure Cys and (b) CAE-AgNPs in the region 200-3000 cm$^{-1}$ after the addition of Cys.

Notably indicating the extent of bonding of Cys with AgNPs surface via carboxyl groups present in it[7,38,39]. Some previous theoretical findings suggested the fact that the amino group in Cys will be weakened due to the dominant nature of interconnecting H-bonds in the case of adsorption on the surface of CAE-AgNPs[38]. To find out the detection limit of our biosensor, Raman study of different dilutions of aqueous Cys solution using CAE-AgNPs (A1) as a substrate was done. Fig. 8 demonstrates the Raman spectra of a mixture of CAE-AgNP colloids with different dilutions of Cys ranging from 0.1 nM to 1 mM. However, Raman signals are observed even at a concentration of 0.1 nM, which indicates, that using our CAE-AgNPs as a substrate the lower detection limit for Cys can reach up to 0.1 nM, or even lower. Hence, it is ensured that our green synthesized AgNPs using CAE can be employed as an effective biosensor for rapid and direct Raman detection of Cys, which can be applied to the biomedical and food industry.

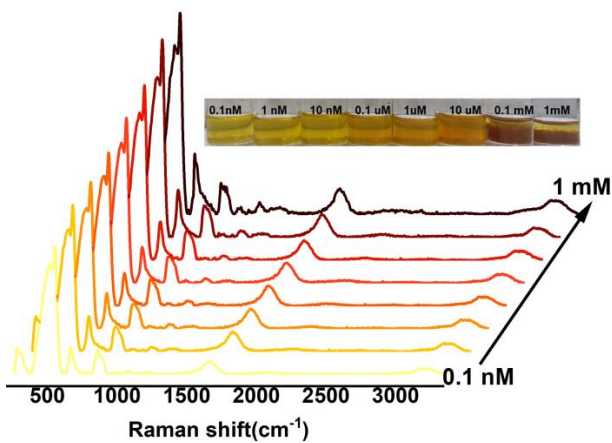

**Figure 8.** Raman spectra and photographs of CAE-AgNPs after the addition of Cys at different concentrations.

**Conclusions**

To the best of our knowledge, this work is the first attempt to demonstrate a highly stable and ecological friendly phytosynthesized AgNPs using CAE. The prepared AgNPs having an average diameter of 25 nm showed excellent stability with a zeta potential value around -57 mv after 3 months from synthesis. We have revealed the reduction mechanism behind the formation of NPs by the dominant alkaloid caffeine as the sole stabilizer and reducing agent with evident support from characteristic peaks and corresponding changes that occurred in spectroscopic studies. The zero-valent and spherical AgNPs with a diameter range of 28 nm were tightly associated with CAE, forming a stable CAE−AgNPs colloid suspension ready for colorimetric detection. The selective detection of Cys containing a thiol-group among 9 amino acids has been confirmed by monitoring the color change from yellow to brown-red with naked eyes and UV−Vis spectroscopy in just 5 min. We performed a detailed study on the chemical interaction after the adsorption of Cys on the surface of CAE-AgNPs using conventional Raman spectroscopy that showed a lower

LOD even at 0.1 nM. Therefore, this study provides a new method for the synthesis of AgNPs that can detect Cys harmlessly, rapidly, sensitively, and selectively.

**Author information**

*Corresponding author*

Email: goutam@nitc.ac.in

*Notes*

The authors have no conflict of interest to declare.

**Acknowledgements**

We thank to Centre for Materials Characterization Facility, NIT Calicut for the Confocal Raman Spectrometer measurements. We thank the departments of chemistry, chemical engineering and the school of material science of the National Institute of Technology, Calicut, India for the necessary instrumental facilities. This work was supported by the Faculty Research Grant [NITC/DEAN(R&C)/FRG/2018- 19/3], NIT Calicut, Kozhikode - 673601, Kerala, India.

**Figure Caption:**

**Scheme 1.** Schematic synthesis procedure for CAE-AgNPs

**Figure 1**. UV-Vis spectrum for (a) aqueous solution of CAE (b) CAE-AgNPs formed using various $AgNO_3$ concentrations; (c) CAE-AgNPs formed using various CAE volumes. (d) DLS spectrum of CAE-AgNPs.

**Figure 2**. Proposed binding scheme of the major phytochemicals of CAE (a) caffeine and (b) CGA.

**Figure 3.** TEM images of CAE-AgNPs at different magnifications (a-c), lattice fringe (d), SAED pattern (e) and histogram for size distribution (f).

**Figure 4**. FTIR-ATR spectra in the region 4000-400 $cm^{-1}$ of CAE (40mg/ml) and CAE-AgNPs (3 mM of $AgNO_3$ with 1ml CAE) are indicated by black and red solid lines, respectively.

**Figure 5.** Raman spectra of (a) CAE (40mg/ml) and (b) CAE-AgNPs (3 mM of $AgNO_3$ with 1ml CAE) in the region of 200-1800 $cm^{-1}$.

**Figure 6.** UV-Vis spectra of (a) selectivity study using different amino acids and the inset shows color noticed for CAE-Ag colloid after the addition of nine amino acids (I- Ala, II- Ba, III- Gly, IV- Cys, V- Arg, VI-Trp, VII- Ga, VIII-Val and IX- Ils), respectively. (b) sensitivity study of Cys at different dilutions and (c) - (d) TEM images of CAE-AgNPs after adding Cys.

**Figure 7.** Raman spectra of (a) pure Cys and (b) CAE-AgNPs in the region 200-3000 $cm^{-1}$ after the addition of Cys.

**Figure 8.** Raman spectra and photographs of CAE-AgNPs after the addition of Cys at different concentrations.

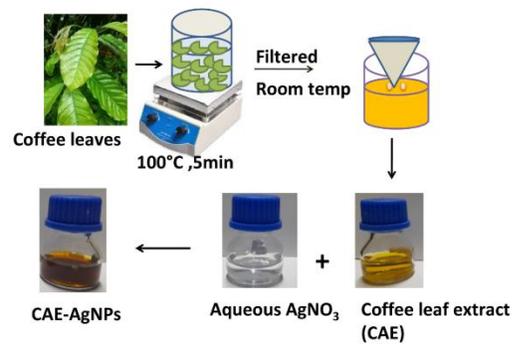

**Scheme 1.**

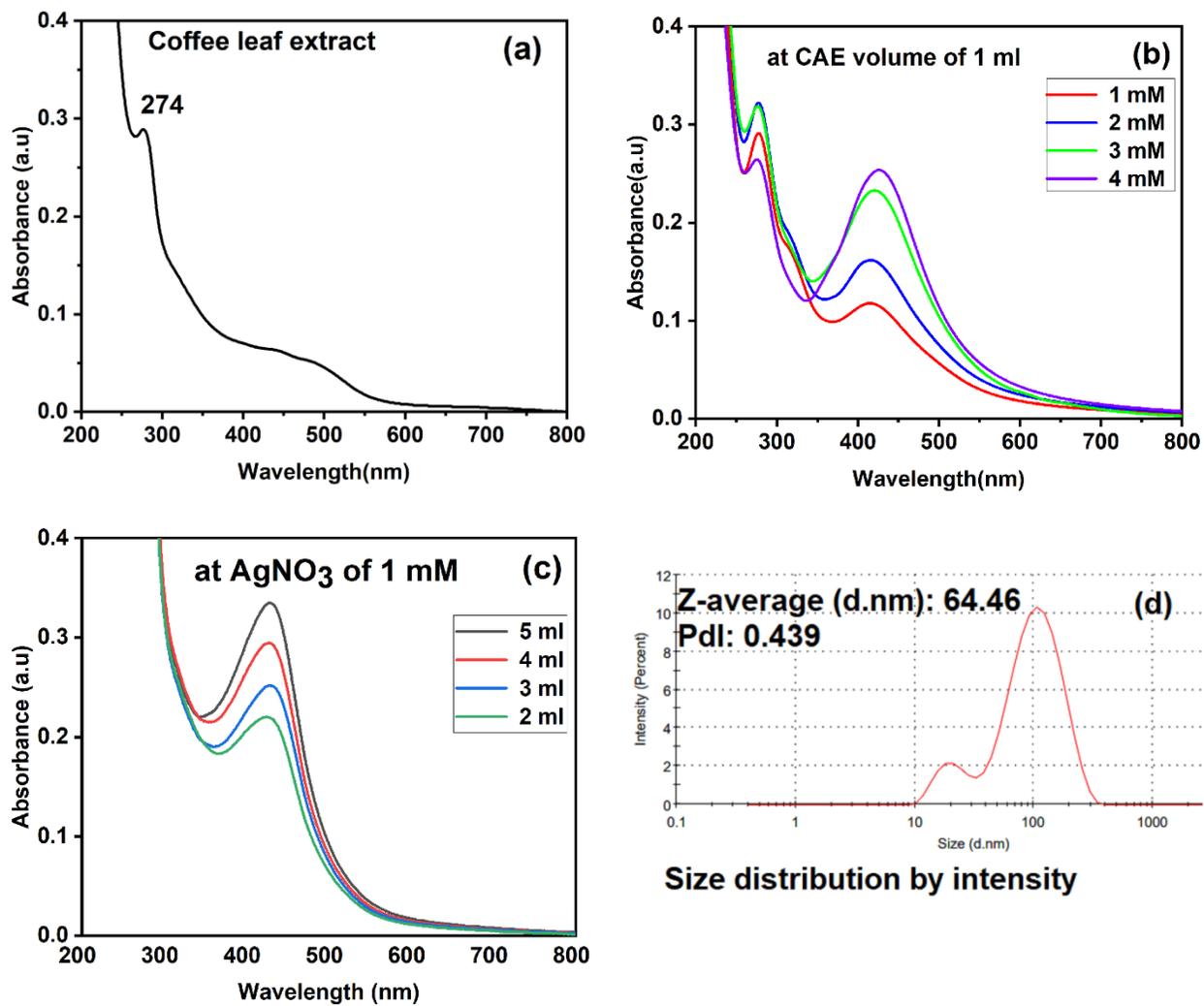

**Figure 1**.

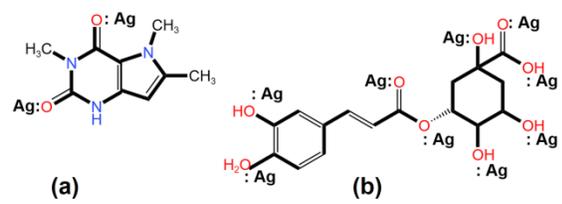

**Figure 2.**

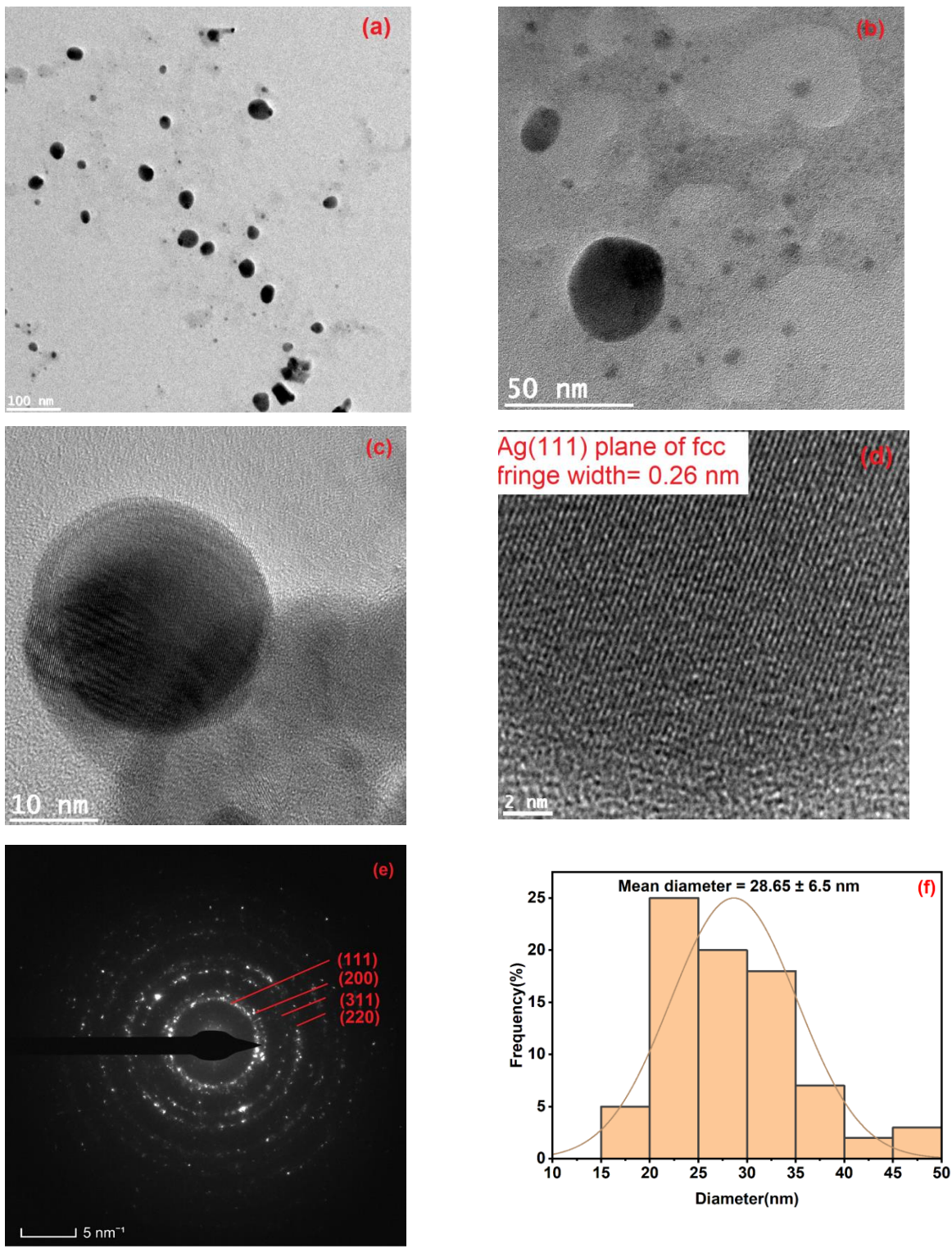

**Figure 3.**

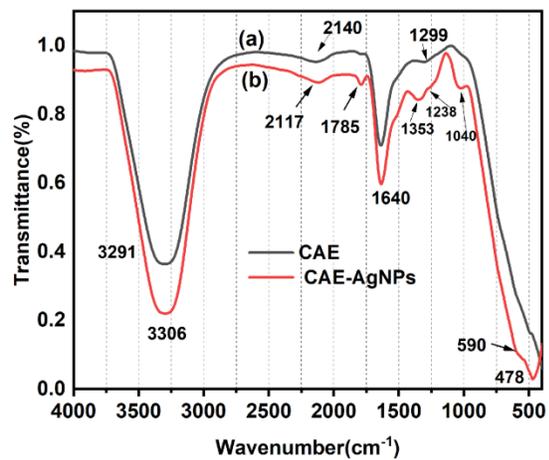

**Figure 4.**

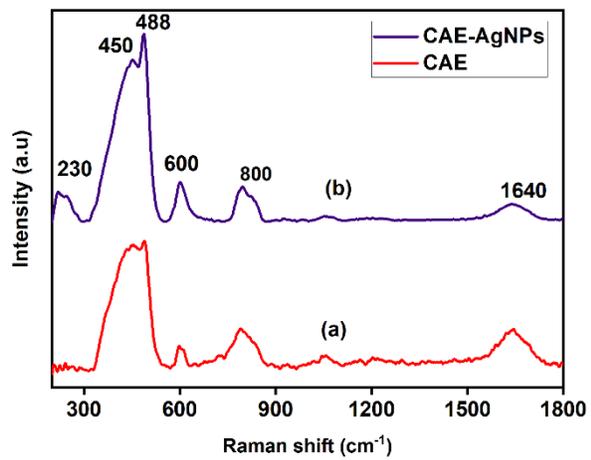

**Figure 5.**

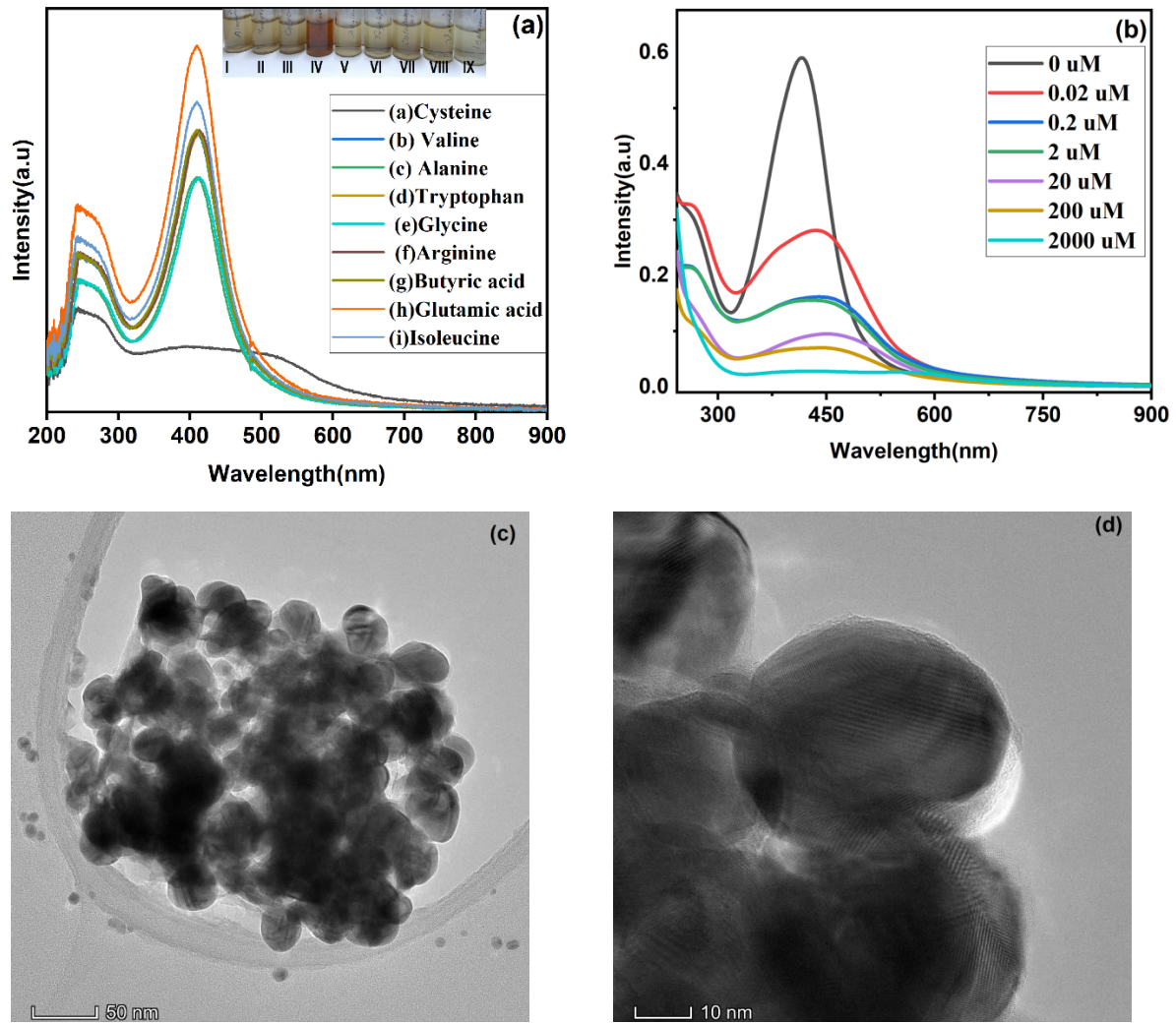

**Figure.6**

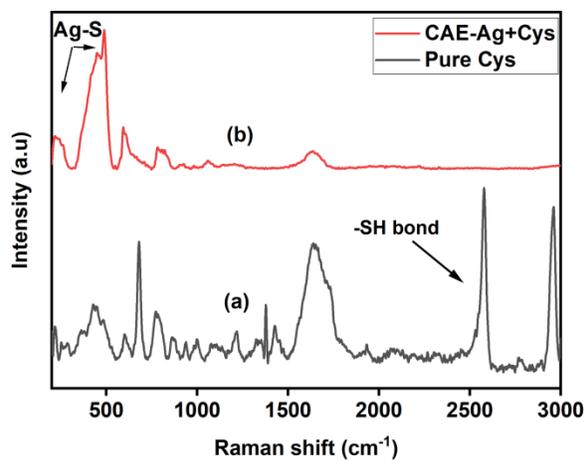

**Figure 7.**

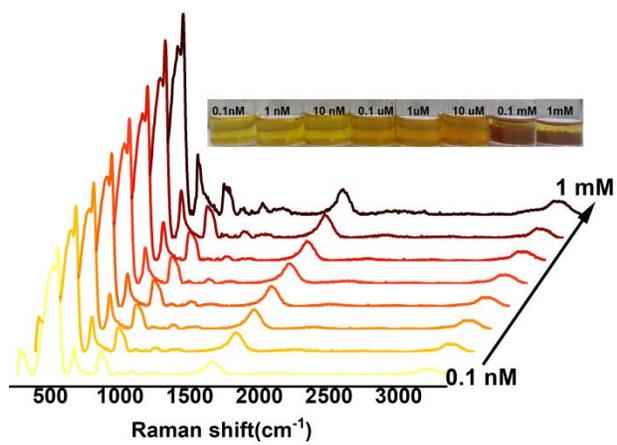

**Figure 8.**

**Table Caption:**

**Table 1.** FTIR vibrational mode assignment for CAE and CAE-AgNPs.

**Table 2**. Raman vibrational mode assignment for CAE and CAE-AgNPs.

**Table 1.**

| Sample | Wavenumber (cm$^{-1}$) | IR Assignments | Reference |
|---|---|---|---|
| CAE | 1299 | Vibrations of polyols | 37 |
| | 1640 | C=O stretching vibration | 34,37 |
| | 2140 | C-H vibrations | 34 |
| | 3291 | O-H stretch | 35,37 |
| CAE-AgNPs | 478, 590 | Ag-O, Ag-N vibrations | 35,37 |
| | 1040, 1238 | Vibrations of primary alcohol (-OH) and polyols | 34, 37 |
| | 1353 | Aromatic band vibrations | 35 |
| | 1640 | C=O, C=N stretch | 34 |
| | 3306 | O-H stretch | 35,37 |

**Table 2.**

| Sample | Wavenumber (cm$^{-1}$) | Raman Assignments | Reference |
|---|---|---|---|
| CAE | 601, 787 | C=C vibrations of alkenes | 34,36 |
| | 1640 | C=O stretching vibration | 34 |
| CAE-AgNPs | 230, 450, 488 | Ag-O, Ag-N vibrations | 37 |
| | 600, 800 | C=C vibrations of alkenes | 34, 36 |
| | 1353 | Aromatic band vibrations | 35 |
| | 1640 | C=O, C=N stretch | 34 |

**Table of contents**

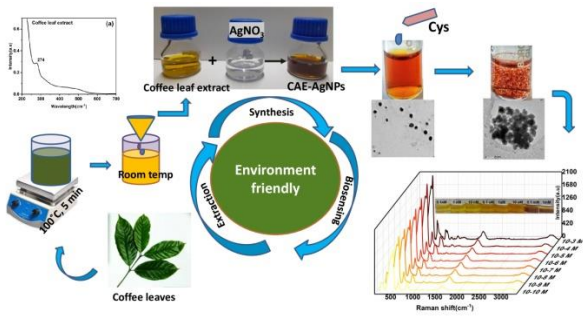